\documentstyle[prd,aps]{revtex}
\begin{document}
\draft
\date{September 15, 2001}
\title{Off--Diagonal Metrics and Anisotropic Brane Inflation}

\author{Sergiu I. Vacaru$^{1}$ and Denis Gontsa$^{2}$}

\address{~}

\address{$^1$Dept. of Phys., CSU Fresno, 2345 East San Ramon Ave.
M/S 37 Fresno, CA 93740-8031, USA \\
 sergiuvacaru@venus.nipne.ro, sergiu$_{-}$vacaru@yahoo.com}

\address{~}

\address{$^2$Department of Physics, St. Petersbourg State University,\\
 P. O. Box 122, Peterhoff, St. Petersbourg, 198904, Russia \\
 d$_{-}$gontsa@yahoo.com}

\maketitle

\begin{abstract}
We study anisotropic reheating (entropy production) on 3D brane from a
decaying bulk scalar field in the brane--world picture of the Universe by
considering off--diagonal metrics and anholonomic frames. We show that a
significant amount of, in general, anisotropic dark radiation is produced in
this process unless only the modes which satisfy a specific relation are
excited. We conclude that subsequent entropy production within the brane is
required in general before primordial nucleosynthesis and that the presence
of off--diagonal components (like in the Salam, Strathee and Petracci works
\cite{salam}) of the bulk metric modifies the processes of entropy
production which could substantially change the brane--world picture of the
Universe.
\end{abstract}


The brane world picture of the Universe \cite{rs} resulted in a number of
works on brane world cosmology \cite{wc1,wc2} and inflationary solutions and
scenaria \cite{inf1,inf2,inf3,inf4,inf5}. Such solutions have been
constructed by using diagonal cosmological metrics with respect to holonomic
coordinate frames.

In Kaluza--Klein gravity there were also used off--diagonal five dimensional
(in brief, 5D) metrics beginning Salam and Strathee and Perrachi works \cite
{salam} which suggested to treat the off--diagonal components as some
coefficients including $U(1), SU(2)$ and $SU(3)$ gauge fields. Recently, the
off--diagonal metrics were considered in a new fashion both in Einstein and
brane gravity \cite{v1,v2}, by applying the method of anholonomic frames
with associated nonlinear connection, which resulted in a new method of
construction of exact solutions of Einstein equations describing, for
instance, static black hole and cosmological solutions with ellipsoidal or
torus symmetry, soliton--dilaton and wormhole--flux tube configurations with
anisotropic polarizations and/or running of constants.

The aim of this paper is to investigate reheating after
anisotropic inflation in the brane world with generic local
anisotropy induced by off--diagonal metrics in the bulk
\cite{v2}. In this scenario, our locally anisotropic Universe is
described on a 4D boundary (3D anisotropic brane) of
$Z_2$--symmetric 5D space--time with a gravitational vacuum
polarization constant and its computed renormalized effective
value. In the locally isotropic limit the constant of
gravitational vacuum polarization results in a negative
cosmological constant $\Lambda_5\equiv -6k^2$, where $k$ is a
positive constant. Our approach is in the spirit of
Horava--Witten theory \cite{horava,lukas} and recovers the
Einstein gravity around the brane with positive tension
\cite{rs,tanaka,sms}, the considerations being extended with
respect to anholonomic frames.

Theories of gravity and/or high energy physics must satisfy a
number of cosmological tests including cosmological inflation
\cite{inf} which for brane models could directed by anisotropic
renormalizations of parameters \cite{v2}. We shall develop a
model of anisotropic inflation scenarios satisfying the next three
requirements:\ 1) it is characterized by a sufficiently long
quasi--exponential expansion driven by vacuum--like energy
density of the potential energy of a scalar field;\ 2) the
termination of accelerated anisotropic expansion is associated
with an entropy production or reheating to stisfy the conditions
for the initial state of the classical hot Big Bang cosmology,
slightly anisotropically deformed, before the primordial
nucleosynthesis \cite{ns} and 3)\ generation of primordial
fluctuations with desired amplitude and spectrum \cite{fluc}.

We assume the 5D vacuum Einstein equations written with respect to
anholonomic frames which for diagonal metrics with respect to holonomic
frames contains a negative cosmological constant $\Lambda _5$ and a 3D brane
at the 5th coordinate $w=0$ about which the space--time is $Z_2$ symmetric
and consider a quadratic line interval
\begin{equation}
\delta s^2=\Omega ^2(t,w,y)[dx^2+g_2\left( t,w\right) dt^2+g_3\left(
t,w\right) dw^2+h_4(t,w,y)\delta y^2+h_5\left( t,w\right) dz^2],
\label{dmetr1}
\end{equation}
where the 'elongated' differential
 $\delta y=dy+\zeta _2(t,w,y)dt+\zeta _3(t,w,y)dw,$
 together with $dx,dt$ and $dw$ define an anholonomic co--frame basis $%
(dx,dt,dw,\delta y,$ $\delta z=dz)$ which is dual to the anholonomic frame
basis \cite{v1,v2} $
(\delta _1=\frac \partial {\partial x},\delta _2=\frac \partial
{\partial t}-\zeta _2\frac \partial {\partial y},\delta _3=\frac \partial
{\partial t}-\zeta _3\frac \partial {\partial y},\partial _4=\frac \partial
{\partial y},\partial _5=\frac \partial {\partial z});$\
we denote the 4D space--time coordinates as $(x,t,w,y,z)$ with $t$ being the
time like variable. The metric ansatz for the interval (\ref{dmetr1}) is
off--diagonal with respect to the usual coordinate basis $\left(
dx,dt,dw,dy,dz\right).$

As a particular case we can parametrize from (\ref{dmetr1}) the metric near
an locally isotropic brane like a flat Robertson--Walker metric with the
scale factor $a(t)$ \cite{inf5} if we state the values
\[
\Omega ^2=(aQ)^2,g_2=-(aQ)^{-2}N^2,g_3=(aQ)^{-2},h_4=1,h_5=1,\zeta _{2,3}=0,
\]
for
\begin{eqnarray}
N^2(t,w) &=&Q^{-2}(t,w)\left[ \cosh (2kw)+{\frac 12}k^{-2}\left( {H^2+\dot{H}%
}\right) \left( {\cosh (2kw)-1}\right) -{\frac{{1+{\frac 12}k^{-2}\left( {%
2H^2+\dot{H}}\right) }}{{\sqrt{1+k^{-2}H^2+Ca^{-4}}}}}\sinh (2k|w|)\right]
\nonumber \\
Q^2(t,w) &=&\cosh (2kw)+{\frac 12}k^{-2}H^2\left( {\cosh (2kw)-1}\right) -%
\sqrt{1+k^{-2}H^2+Ca^{-4}},  \label{NQ}
\end{eqnarray}
when the bulk is in a vacuum state with a negative cosmological constant $%
\Lambda _5$, $C$ is an integration constant \cite{mukoyama}. One takes $%
N=Q=1 $ on the brane $w=0$. The function $H\left( t\right) $ and constants $%
k $ and $C$ from (\ref{NQ}) are related with the evolution equation on the
brane in this case is given by
\begin{equation}
H^2 = \left( \frac{\dot{a}}a\right) ^2={\frac{{\kappa _5^4\sigma }}{{18}}}%
\rho _{{\rm tot}}+{\frac{{\Lambda _4}}3}+{\frac{{\kappa _5^4}}{{36}}}\rho _{%
{\rm tot}}^2-{\frac{{k^2C}}{{a^4}}},\
\Lambda _4 \equiv {\frac 12}\left( {\Lambda _5+{\frac{{\kappa _5^2}}6}%
\sigma ^2}\right) ,  \label{feq}
\end{equation}
where $\kappa _5^2$ is the 5D gravitational constant related with the 5D
reduced Planck scale, $M_5$, by $\kappa _5^2=M_5^{-3}$; $\sigma $ is the
brane tension, the total energy density on the brane is denoted by $\rho _{%
{\rm tot}}$, and the last term of (\ref{feq}) represents the dark radiation
with $C$ being an integration constant \cite{sms,bki,mukoyama}. We recover
the standard Friedmann equation with a vanishing cosmological constant at
low energy scales if $\sigma =6k/\kappa _5^2$ and $\kappa _4^2=\kappa
_5^4\sigma /6=\kappa _5^2k$, where $\kappa _4^2$ is the 4D gravitational
constant related with the 4D reduced Planck scale, $M_4$, as $\kappa
_4^2=M_4^{-2}$. We find that $M_4^2=M_5^3/k$. If we take $k=M_4$, all the
fundamental scales in the theory take the same value, i. e. $k=M_4=M_5.$ The
the scale above is stated by constant $k$ which the nonstandard term
quadratic in $\rho _{{\rm tot}}$ is effective in (\ref{feq}). One suppose
\cite{inf5} that $k$ is much larger than the scale of inflation so that such
quadratic corrections are negligible.

For simplicity, in locally isotropic cases one assumes that the bulk metric
is governed by $\Lambda _5$ and neglect terms suppressed by $k^{-1}$ and $%
Ca^{-4}$ and writes
\begin{equation}
ds_5^2=-e^{-2k|w|}dt^2+e^{-2k|w|}a^2(t)\left( {dx^2+dy^2+dz^2}\right) +dw^2.
\label{met}
\end{equation}

The conclusion of Refs \cite{v2} is that the presence of off--diagonal
components in the bulk 5D metric results in locally anisotropic
renormalizations of fundamental constants and modification of the Newton low
on the brane. The purpose of this paper is to analyze the basic properties
of models of anisotropic inflation on 3D brane with induced from the bulk
local anisotropy of metrics of type (\ref{dmetr1}) which define cosmological
solutions of 5D vacuum Einstein equations depending on variables $\left(
w,t,y\right) ,$ being anisotropic on coordinate $y$ (see details on
construction of various classes of solutions by applying the method of
moving anholonomic frames in Ref \cite{v1,v2}).

For simplicity, we shall develop a model of inflation on 3D brane with
induced from the bulk local anisotropy by considering the ansatz
\begin{equation}
\delta s_5^2 =e^{-2(k|w|+k_y|y|)}a^2(t)[dx^2-dt^2]+e^{-2k_y|y|}dw^2+
e^{-2(k|w|+k_y|y|)}a^2(t)({\delta y}^2+{z^2)}  \label{meta}
\end{equation}
which is a particulcar case of the metric (\ref{dmetr1}) with\
$\Omega ^2(w,t,y)=e^{-2(k|w|+k_y|y|)}a^2(t),{g}_2=-1,
g_3=e^{2k|w|},h_4=1,h_5=1$
and ${\zeta }_2=k/k_y,{\zeta }_3=(da/dt)/k_ya$ taken as the ansatz (\ref
{meta}) would be an exact solution of 5D vacuum Einstein equations. The
constants $k$ and $k_y$ have to be established experimentally. We emphasize
that the metric (\ref{meta}) is induced alternatively on the brane from the
5D anholonomic gravitational vacuum with off--diagonal metrics. With respect
to anholonomic frames it has some diagonal coefficients being similar to
those from (\ref{met}) but these metrics are very different in nature and
describes two types of branes: the first one is with generic off--diagonal
metrics and induced local anisotropy, the second one is locally isotropic
defined by a brane configuration and the bulk cosmological constant. For
anisotropic models, the respective constants can be treated as some
'receptivities' of the bulk gravitational vacuum polarization.

The next step is to investigate a scenarios of anisotropic inflantion driven
by a bulk scalar field $\phi $ with a 4D potential $V[\phi ]$ \cite{bbinf,hs}%
. We shall study the evolution of $\phi $ after anisotropic brane
inflation expecting that reheating is to proceed in the same way
as in 4D theory with anholonomic modification (a similar idea is
proposed in Ref. \cite{bas} but for locally isotropic branes). We
suppose that the scalar field is homogenized in 3D space as a
result of inflation, it depends only on $t$ and $w$ and
anisotropically on $y$ and consider a situation when $\phi $
rapidly oscillates around $\phi =0$ by expressing $V[\phi
]=m^2\phi ^2/2$. The field $\phi (t,w,y)$ is non--homogeneous
because of induced space--time anisotropy. Under such assumptions
he Klein--Gordon equation in the background of metric
(\ref{meta}) is written
\begin{equation}
\Box _5\phi (f,t,y)-V^{\prime }[\phi (f,t,y)] =\frac 1{\sqrt{|g|}}\left[
\delta _t\left( \sqrt{|g|}g^{22}\delta _t\phi \right) +\delta _w\left( \sqrt{%
|g|}g^{33}\delta _f\phi \right) +\partial _y\left( \sqrt{|g|}h^{44}\partial
_y\phi \right) \right] -V^{\prime }[\phi ]=0,  \label{kg}
\end{equation}
where $\delta _w =\frac \partial {\partial w}-
\zeta _2\frac \partial {\partial y},\delta _t=\frac \partial {\partial t}-\zeta _3\frac \partial {\partial y},
 $, $\Box _5$ is the d'Alambert operator and $|g|$ is the determinant of
the matrix of coefficients of metric given with respect to the anholonomic
frame (in Ref. \cite{inf5} the operator $\Box _5$ is alternatively
constructed by using the metric (\ref{met})).

The energy release of $\phi $ is modeled by introducing
phenomenologically a
dissipation terms defined by some constants $\Gamma _D^w,\Gamma _D^y$ and $%
\Gamma _B$ representing the energy release to the brane and to the entire
space,
\begin{equation}
\Box _5\phi (w,t,y)-V^{\prime }[\phi (w,t,y)]=\frac{\Gamma _D^w}{2k}\delta
(w)\frac 1N\delta _t\phi +\frac{\Gamma _D^y}{2k_y}\delta (y)\frac 1N\delta
_t\phi +\Gamma _B\frac 1N\delta _t\phi .  \label{kgv}
\end{equation}

Following (\ref{kg}) and (\ref{kgv}) together with the $Z_2$ symmetries on
coordinates $w$ and $y$, we have
\begin{equation}
\delta _w\phi ^{+}=-\delta _w\phi ^{-}=\frac{\Gamma _D^w}{4k}\delta _t\phi
(0,y,t),\partial _y\phi ^{+}=-\partial _y\phi ^{-}=\frac{\Gamma _D^y}{4k_y}%
\delta _t\phi (w,0,t),  \label{phiw}
\end{equation}
where superscripts $+$ and $-$ imply values at $w,y\longrightarrow +0$ and $%
-0$, respectively. In this model we have two types of warping
factors, on coordinates $w$ and $y.$ The constant $k_y$
characterize the gravitational anisotropic polarization in the
direction $y.$

Comparing the formulas (\ref{kgv})$\,$and (\ref{phiw}) with similar ones
from Ref. \cite{inf5} we conclude the the induced from the bulk brane
anisotropy could result in additional dissipation terms like that
proportional to $\Gamma _D^y$. This modifies the divergence of divergence $%
T_{~~A;C}^{(\phi )C}$ of the energy--momentum tensor $T_{MN}^{(\phi )}$ of
the scalar field $\phi $: Taking
\[
T_{MN}^{(\phi )}={{\delta }_M}\phi {{\delta }_N}\phi -g_{MN}\left( {{\frac 12%
}g^{PQ}{\delta }_P\phi {\delta }_Q\phi +V[\phi ]}\right) ,
\]
with five dimensional indices, $M,N,...=1,2,...,5$ and anholonomic partial
derivative operators ${{\delta }_P}$ being dual to $\delta ^P$ we compute
\begin{equation}
T_{~~A;C}^{(\phi )C}=\left\{ \Box _5\phi (w,t,y)-V^{\prime }[\phi
(w,t,y)]\right\} \phi _{,A}=\left[ \frac{\Gamma _D^w}{2k}\delta (w)\frac
1N\delta _t\phi +\frac{\Gamma _D^y}{2k_y}\delta (y)\frac 1N\delta _t\phi
+\Gamma _B\frac 1N\delta _t\phi \right] \delta _A\phi .  \label{divergence}
\end{equation}
We can integrate the $A=0$ component of (\ref{divergence}) from
$w=-\epsilon $ to $w=+\epsilon $ near the brane, than we
integrate from $y=-\epsilon _1$ to $y=+\epsilon _1,$ in the zero
order in $\epsilon $ and $\epsilon _1,$ we find from (\ref{phiw})
that
\begin{equation}
\frac{\delta \rho _\phi (0,0,t)}{\partial t}=-(3H+\Gamma _B)(\delta _t\phi
)^2(0,0,t)-J_\phi (0,0,t),  \label{rhophieq}
\end{equation}
with
$\rho _\phi \equiv \frac 12(\delta _t\phi )^2+V[\phi ],~~J_\phi \equiv -%
\frac{\delta _t\phi }{\sqrt{|g|}}\delta _f\left( \sqrt{|g|}\delta _f\phi
\right) -\frac{\delta _t\phi }{\sqrt{|g|}}\delta _y\left( \sqrt{|g|}\delta
_y\phi \right),$
which states that the energy dissipated by the $\Gamma _D^f$ and $\Gamma _D^y
$ terms on anisotropic brane is entirely compensated by the energy flows
(locally isotropic and anisotropic) onto the brane. In this paper we shall
model anisotropic inflation by considering that $\phi $ looks like
homogeneous with respect to anholonomic frames; the local anisotropy and
induced non--homogeneous effects are modeled by additional terms like $%
\Gamma _D^y$ and elongated partial operators with a further integration on
variable $y.$

Now we analyze  how both the isotropic and anisotropic energy released from $%
\phi $ affects evolution of our brane Universe by analyzing gravitational
field equations \cite{sms,hs} written with respect to anholonomic frames. We
consider that the total energy--momentum tensor has a similar structure as
in holonomic coordnates but with the some anholonomic variables, including
the contribution of bulk cosmological constant,
\[
T_{MN}=-\kappa _5^{-2}\Lambda _5g_{MN}+T_{MN}^{(\phi )}+S_{MN}\delta (w),
\]
where $S_{MN}$ is the stress tensor on the brane and the capital
Latin indices $M,N,...$ run values $1,2,...5$ (we follow the
denotations from \cite {inf5} with that difference that the
coordinates are reordered and stated with respect to anholonomic
frames). One introduces a  further decomposition as $S_{\mu \nu
}=-\sigma q_{\mu \nu }+\tau _{\mu \nu },$ where $\tau _{\mu \nu
}$ represents the energy--momentum tensor of the radiation fields
produced by the decay of $\phi $ and it is of the form $\tau _\nu
^\mu ={\rm diag}(2p_r,-\rho _r,p_r,0)$ with $p_r=\rho _r/3$ which
defines an anisotropic distribution of matter because of
anholonomy of the frame of reference.

We can remove the considerations on an anisotropic brane (hypersurface) by
using a unit vector $n_M=(0,0,1,0,0)$ normal to the brane for which the
extrinsic curvature of a $w=const$ hypersurface is given by $%
K_{MN}=q_M^Pq_N^Qn_{Q;P}$ with $q_{MN}=g_{MN}-n_Mn_N$. Applying the Codazzi
equation and the 5D Einstein equations with anholonomic variables \cite
{v1,v2}, we find
\begin{equation}
D_\nu K_\mu ^\nu -D_\mu K=\kappa _5^2T_{MN}n^Nq_\mu ^M=\kappa _5^2T_{\mu
w}=\kappa _5^2(\delta _t\phi )(\delta _w\phi )\delta _\mu ^2,  \label{k1}
\end{equation}
where $\delta _\mu ^1$ is the Kronecker symbol, small Greek indices
parametrize coodinates on the brane, $D_\nu $ is the 4D covariant derivative
with respect to the metric $q_{\mu \nu }$. The above equation reads
\begin{equation}
D_\nu K_0^{\nu +}-D_0K^{+}=\kappa _5^2\left[ {\frac{\Gamma _D}{{4k}}(}\delta
_t\phi )^2(0,t,0)+{\frac{\Gamma _D}{{4}k_y}(}\partial _y\phi
)^2(0,t,0)\right] ,  \label{k1a}
\end{equation}
near the brane  $w\longrightarrow +0$ and neglecting
non--homogeneous behaviour, by putting $y=0.$ We have
\begin{equation}
D_\nu K_\mu ^{\nu +}-D_\mu K^{+}=-{\frac{{\kappa _5^2}}2}D_\nu S_\mu ^\nu =-{%
\frac{{\kappa _5^2}}2}D_\nu \tau _\mu ^\nu .  \label{eqa1}
\end{equation}
which follows from the junction condition and  $Z_2$--symmetry with $K_{\mu
\nu }^{+}=-{\frac{{\kappa _5^2}}2}\left( {S_{\mu \nu }-{\frac 13}q_{\mu \nu
}S}\right) .$ Using (\ref{k1a}) and (\ref{eqa1}), we get
\[
D_\nu \tau _\mu ^\nu =-{\frac{\Gamma _D}{2k}(}\delta _t\phi )^2\delta _\mu
^2-{\frac{\Gamma _D^y}{2k_y}(}\partial _y\phi )^2\delta _\mu ^3,
\]
i. e.,
\[
\delta _t{\rho _r}=-3H(\rho _r+p_r)+{\frac{\Gamma _D}{2k}(}\delta _t\phi )^2+%
{\frac{\Gamma _D^y}{2k_y}(}\partial _y\phi )^2=-4H\rho _r+{\frac{\Gamma _D}{%
2k}(}\delta _t\phi )^2+{\frac{\Gamma _D^y}{2k_y}(}\partial _y\phi )^2,
\]
on the anisotropic brane. This equation describe the reheating in an
anisotropic  perturbation theory (for  inflation in 4D theory see \cite{prt}%
).

The 4D Einstein equations with the Einstein tensor  $G_{~\mu }^{(4)\nu }$
were proven  \cite{hs} to have the form
\[
G_{~\mu }^{(4)\nu }=\kappa _4^2\left( {T_{~\mu }^{(s)\nu }+\tau _\mu ^\nu }%
\right) +\kappa _5^4\pi _\mu ^\nu -E_\mu ^\nu ,
\]
with $
T_{~\mu }^{(s)\nu }\equiv {\frac 1{{6k}}}\left[ {4q^{\nu \zeta }(\delta }%
_\mu {\phi ){(\delta }_\zeta {\phi )}+\left( {{\frac 32{(\delta }_\zeta {%
\phi )}}^2-{\frac 52}q^{\xi \zeta }{(\delta }_\xi {\phi )(\delta }_\zeta {%
\phi )}-\frac 32m^2\phi ^2}\right) q_\mu ^\nu }\right], $
 where  $\pi _\mu ^\nu $ contains
  terms quadratic in $\tau _\alpha ^\beta $ which are higher order in
 $\rho_r/(kM_4)^2$ and are consistently neglected in our analysis.
$E_\mu ^\nu \equiv C_{\mu w}^{w\nu }$ is a component of the 5D Weyl
 tensor $C_{PQ}^{MN}$ treated as a source of  dark radiation \cite{mukoyama}.

With respect to anholonomic frames the 4D Bianchi identities are written in
the usual manner,
\begin{equation}
D_\nu G_{~\mu }^{(4)\nu }=\kappa _4^2\left( {D_\nu T_{~\mu }^{(s)\nu }+D_\nu
\tau _\mu ^\nu }\right) -D_\nu E_\mu ^\nu =0,  \label{bianchi}
\end{equation}
with that difference that $D_\nu G_{~\mu }^{(4)\nu }=0$ only for holonomic
frames but in the anholonomic cases, for general constraints one could be
 $D_\nu G_{~\mu }^{(4)\nu }\neq 0$ \cite{v1,v2}. In this paper we shall
consider such constraints for which the equalities (\ref{bianchi}) hold
which  yield
\[
D_\nu E_2^\nu =-{\frac{{\kappa _4^2}}{{4k}}}{\frac \delta {{\partial t}}}%
\left[ {{(}\delta _t\phi )^2-{(}\delta _w\phi )^2-{(\partial }}_y{\phi
)^2+m^2\phi ^2}\right] -\frac{2\kappa _4^2H}k{{(}\delta _t\phi )^2}-{\frac{{%
\kappa _4^2}}{2k}}\Gamma _D{{(}\delta _t\phi )^2}-{\frac{{\kappa _4^2}}{2k}}%
_y\Gamma _D^y{(\partial }_y{\phi )^2}.
\]
Putting on the anisotropic brane  ${{(}\delta _w\phi )^2=}\Gamma _D^2{{(}%
\delta _t\phi )^2}/(16k^2)$ and ${(\partial }_y{\phi )^2=(}\Gamma _D^y)^2{{(}%
\delta _t\phi )^2}/(16k_y^2)$ similarly to Ref. \cite{inf5}, by substituting
usual partial derivatives into 'elongated' ones and introducing $\varphi
(t)\equiv \phi (0,t.y)/\sqrt{2k},$ $b=a(t)e^{-k_y|y|}$ and $\varepsilon
\equiv E_2^2/\kappa _4^2$ we prove the evolution equations in the brane
universe $w=0$
\begin{eqnarray*}
H^2 &=&\left( {{\frac{\delta _tb}b}}\right) ^2={\frac{{\kappa _4^2}}3}\left(
\rho _\varphi +\rho _r+\varepsilon \right) ,~~~\rho _\varphi \equiv {{\frac
12(}\delta _t\varphi )^2+\frac 12m^2\varphi ^2=\frac{\rho _\phi }{2k}},~~~ \\
{\delta _t}\rho _\varphi  &=&-(3H+\Gamma _B){{(}\delta _t\varphi )^2}%
-J_\varphi ,~~~J_\varphi \equiv \frac{J_\phi }{2k}, \\
{\delta _t\rho _r} &=&-4H\rho _r+\Gamma _D{{(}\delta _t\varphi )^2}+\Gamma
_D^y{{(}\delta _t\varphi )^2}, \\
{\delta _t\varepsilon } &=&-4H\varepsilon -(H+\Gamma _D+\Gamma _D^y-\Gamma
_B){{(}\delta _t\varphi )^2}+J_\varphi .
\end{eqnarray*}

We find the solution of (\ref{kg}) in the background  (\ref{meta}) in the
way suggested by \cite{gw,inf5} by introducing an additional factor
depending on anisotropic variable $y,$
\[
\phi (t,w)=\sum\limits_{n+n^{\prime }}{c_{n+n^{\prime }}T_{n+n^{\prime
}}(t)Y_n(w)Y_{n^{\prime }}(y)}+H.C.
\]
with
\begin{eqnarray*}
T_{n+n^{\prime }}(t) &\cong &a^{-{\frac 32}}(t)e^{-i(m_n+m_{n^{\prime }})t},
\\
Y_n(w) &=&e^{2k|w|}\left[ {J_\nu ^{}\left( {{{\frac{{m_n}}k}}e^{k|w|}}%
\right) +b_nN_\nu ^{}\left( {{{\frac{{m_n}}k}}e^{k|w|}}\right) }\right],\
Y_{n^{\prime }}(y) = e^{2k_y|y|}\left[ {J_\nu ^{}\left( {{{\frac{%
m_{n^{\prime }}}k}}e^{k_y|y|}}\right) +b_{n^{\prime }}N_\nu ^{}\left( {{{%
\frac{{m_n}}k}}e^{k_y|y|}}\right) }\right] ,~
\end{eqnarray*}
for $~\nu =2\sqrt{1+{\frac{{m^2}}{{4k^2}}}}\cong 2+{\frac{{m^2}}{{4k^2}}},$
and considering that the field oscillates rapidly in cosmic expansion time
scale. The values  $m_n$ and $m_{n^{\prime }}$ are some  constants which may
take continuous values in the case of a single brane and $b_n$ and  ${%
b_{n^{\prime }}}$ are some constants determined by the boundary conditions, $%
\delta _w\phi =0$ at $w=0$ and $\partial _y\phi =0$ at $y=0.$ We write
\begin{eqnarray*}
b_n &=&\left[ 2J_\nu \left( \frac{m_n}k\right) +\frac{m_n}kJ_\nu ^{\prime
}\left( \frac{m_n}k\right) \right] \left[ 2N_\nu \left( \frac{m_n}k\right) +%
\frac{m_n}kN_\nu ^{\prime }\left( \frac{m_n}k\right) \right] ^{-1}, \\
b_{n^{\prime }} &=&\left[ 2J_\nu \left( \frac{m_{n^{\prime }}}{k_y}\right) +%
\frac{m_{n^{\prime }}}{k_y}J_\nu ^{\prime }\left( \frac{m_{n^{\prime }}}{k_y}%
\right) \right] \left[ 2N_\nu \left( \frac{m_{n^{\prime }}}{k_y}\right) +%
\frac{m_{n^{\prime }}}{k_y}N_\nu ^{\prime }\left( \frac{m_{n^{\prime }}}{k_y}%
\right) \right] ^{-1}.
\end{eqnarray*}
The effect of dissipation on the boundary conditions is given by
\begin{eqnarray*}
b_n &\cong &\left[ \left( 2+\frac{im_n\Gamma _D}{2k^2}\right) J_\nu \left(
\frac{m_n}k\right) +\frac{m_n}kJ_\nu ^{\prime }\left( \frac{m_n}k\right)
\right] \left[ \left( 2+\frac{im_n\Gamma _D}{2k^2}\right) N_\nu \left( \frac{%
m_n}k\right) +\frac{m_n}kN_\nu ^{\prime }\left( \frac{m_n}k\right) \right]
^{-1}, \\
b_{n^{\prime }} &\cong &\left[ \left( 2+\frac{im_{n^{\prime }}\Gamma _D^y}{%
2k_y^2}\right) J_\nu \left( \frac{m_{n^{\prime }}}{k_y}\right) +\frac{%
m_{n^{\prime }}}{k_y}J_\nu ^{\prime }\left( \frac{m_{n^{\prime }}}{k_y}%
\right) \right] \left[ \left( 2+\frac{im_{n^{\prime }}\Gamma _D^y}{2k_y^2}%
\right) N_\nu \left( \frac{m_{n^{\prime }}}{k_y}\right) +\frac{m_{n^{\prime
}}}{k_y}N_\nu ^{\prime }\left( \frac{m_{n^{\prime }}}{k_y}\right) \right]
^{-1},
\end{eqnarray*}
where use has been made of $\dot{\partial _tT_{n+n^{\prime }}}(t)\cong
-im_{n+n^{\prime }}T_{n+n^{\prime }}(t)$.

For simplicity, let analyze the case when a single oscillation mode exists,
neglect explicit dependence of $\varphi $ on variable $y$ (the effect of
anisotropy being modeled by terms like $m_{n^{\prime }},$ $k_y$ and  $%
\Gamma _D^y$and compare our results with those for isotropic inflation \cite
{inf5}. In this case we approximate $\delta \varphi \simeq \dot{\varphi},$
where dot is used for the partial derivative $\partial _t.$ We find
\begin{equation}
J_\varphi =(m_n^2+m_{n^{\prime }}^2-m^2)\varphi \dot{\varphi}.
\label{jvarphi}
\end{equation}

The evolution of the dark radiation is approximated  in the regime when $%
\varphi (t)$ oscillates rapidly, parametrized as
\[
\varphi (t)=\varphi _i\left( \frac{a(t)}{a(t_i)}\right) ^{-3/2}e^{-i\lambda
_{n+n^{\prime }}(t-t_i)},\ \lambda _{n+n^{\prime }}\equiv m_n+m_{n^{\prime
}}-\frac i2\Gamma _B,
\]
with $m_n+m_{n^{\prime }}\gg H$ and $\Gamma _B$ being positive constants
which  assumes that only a single oscillation mode exists.

Then the evolution equation of $\varepsilon (t)\equiv \kappa _4^{-2}E_2^2$
is given by
\[
\frac{\partial \varepsilon }{\partial t}=-4H\varepsilon -(\Gamma _D+\Gamma
_D^y+H-\Gamma _B)\dot{\varphi}^2-(m^2-m_n^2-m_{n^{\prime }}^2)\varphi \dot{%
\varphi}.
\]
The next approximation is to consider $\varphi $ as oscillating rapidly in
the expansion time scale by averaging the right-hand-side of evolution
equations over an oscillation period. Using $\overline{\dot{\varphi}^2}%
(t)=(m_n^2+m_{n^{\prime }}^2)\overline{\varphi ^2}(t)$ and $\overline{%
\varphi \dot{\varphi}}(t)=-(3H+\Gamma _B)\overline{\varphi
^2}(t)/2$, we obtain the following set of evolution equations in
the anisotropic brane universe $w=0,$ for small non--homogeneities
on $y,$ where the bar denotes average over the oscillation period.
\begin{eqnarray}
H^2 &=&\left( {{\frac{{\dot{a}}}a}}\right) ^2\cong {\frac{{\kappa _4^2}}3}%
\left( \rho _\varphi +\rho _r+\varepsilon \right) ,  \nonumber \\
\frac{\partial \rho _\varphi }{\partial t} &=&-\frac 12(3H+\Gamma
_B)(m^2+m_n^2+m_{n^{\prime }}^2)\overline{\varphi ^2},  \nonumber \\
{\frac{{\partial \rho _r}}{{\partial t}}} &=&-4H\rho _r+\left( \Gamma
_Dm_n^2+\Gamma _D^ym_{n^{\prime }}^2\right) \overline{\varphi ^2},
\label{reqb} \\
{\frac{{\partial \varepsilon }}{{\partial t}}} &=&-4H\varepsilon -(\Gamma
_D+H-\Gamma _B)m_n^2\overline{\varphi ^2}-(\Gamma _D^y+H-\Gamma
_B)m_{n^{\prime }}^2\overline{\varphi ^2}+\frac 12(3H+\Gamma
_B)(m^2-m_n^2-m_{n^{\prime }}^2)\overline{\varphi ^2},  \label{eeqb}
\end{eqnarray}
with $
\overline{\varphi ^2}(t)\equiv \overline{\varphi _i^2}\left( \frac{a(t)}{%
a(t_i)}\right) ^{-3}e^{-\Gamma _B(t-t_i)}.$

The system (\ref{reqb}) and (\ref{eeqb}) was analyzed in Ref. \cite{inf5}
for the case when $m_{n^{\prime }}$ and $\Gamma _D^y$ vanishes:

It was concluded that if  $m_n\geq m,$  we do not recover standard cosmology
on the brane after inflation. The same holds true in the presence of
anisotropic terms.

In the locally isotropic case it was proven that if $m_n\ll n$  the last
term of (\ref{eeqb}) is dominant and we find more dark radiation than
ordinary radiation unless $\Gamma _B$ is extremely small with $\Gamma
_B/\Gamma _D<m_n^2/m^2\ll 1$. The presence of anisotropic values $%
m_{n^{\prime }}$ and $\Gamma _D^y$ can violate this condition.

The cases $m_n,m_{n^{\prime }}\lesssim m$ are  the most delicate
cases because the final amount of dark radiation can be either
positive or negative depending on the details of the model
parameters and the type of anisotropy. The amount of extra
radiation--like matter have to be hardly constrained \cite{ns} if
we wont a successful primordial nucleosynthesis.  In order to have
sufficiently small $\varepsilon $ compared with $\rho _r$ after
reheating without resorting to subsequent both isotropic and
anisotropic entropy production within the brane, the magnitude of
creation
terms of $\varepsilon $ should be vanishingly small at the reheating epoch $%
H\simeq \Gamma _B$. This hold if  only there are presented isotropic and/or
anisotropic modes which  satisfies the inequality
\begin{equation}
\left| 2\Gamma _B(m^2-m_n^2-m_{n^{\prime }}^2)-\Gamma _Dm_n^2-\Gamma
_D^ym_{n^{\prime }}^2\right| \ll \Gamma _Dm_n^2+\Gamma _D^ym_{n^{\prime }}^2.
\label{cond}
\end{equation}

We conclude that the relation (\ref{cond}) should be satisfied for the
graceful exit of anisotropic brane inflation driven by a bulk scalar field $%
\phi $. The presence of off--diagonal components of the metric in the bulk
which induces brane anisotropies could modify the process of
nucleosynthesis.

In summary we have analyzed a model of anisotropic inflation generated
 by a 5D off--diagonal metric in the bulk. We studied entropy production
on the anisotropic 3D brane from a decaying bulk scalar field $\phi $
by considering anholonomic frames and introducing dissipation terms
 to its  equation of motion phenomenologically. We illustrated that
 the dark radiation is significantly produced at the same time unless the
 inequality (\ref{cond}) is satisfied. Comparing our results with
 a similar model in locally isotropic background we found that
 off--diagonal metric components and anisotropy results in additional
 dissipation terms and coefficients which could substantially
modify the scenary of inflation but could not to fall the
qualitative
 isotropic possibilities for well defined cases with specific
 form of isotropic and anisotropic dissipation.
Although we have analyzed only a  case of anisotropic metric with a
specific form of the dissipation, we expect our conclusion is generic and
applicable to other forms of anisotropies and dissipation,
 because it is essentially an outcome of the anholonomic frame method
 and Bianchi identities (\ref{bianchi}). We therefore conclude that
 in the brane--world picture of the Universe it is very important what
 type of metrics and frames we consider, respectively, diagonal
 or off--diagonal and  holonomic or anholonomic (i. e. locally isotropic
 or anisotropic). In all cases there are conditions to be imposed
 on anisotropic parameters and polarizations when
 the dominant part of the entropy we observe experimentally originates
within the brane rather than in the locally anisotropic  bulk.
 Such extra dimensional vacuum gravitational anisotropic
 polarizations of cosmological inflation parameters may be observed
 experimentally.

\vskip 0.5cm 
The authors are grateful to D. Singleton and E. Gaburov for useful
communications. The work of S. V.  is partially supported by the
  "The 2000--2001 California State University Legislative Award".

\end{document}